\documentclass[twocolumn]{aastex631}
\usepackage{multirow}
\usepackage{graphicx}

\newcommand{\Add}[1]{#1}
\newcommand{\Adds}[1]{#1}

\newcommand{\OIII}{{\rm [OIII]\,$\lambda$5007\,}}
\newcommand{\NHagn}{$N_{{\rm H},\Add{agn}}$\,}
\newcommand{\voff}{$\bar{v}_{los}\,$}
\newcommand{\sigv}{$\sigma_{v_{los}}\,$}
\newcommand{\pp}[2]{$P$({#1}$>$ {#2} km s$^{-1}$)}

\usepackage{amsmath}
\shorttitle{AASTeX v6.3.1 Sample article}
\shortauthors{Yutani et al.}

\begin{document}
\title{Apparent effect of dust extinction on the observed outflow velocity of ionized gas in galaxy mergers}

\author{Naomichi Yutani}
\affiliation{Kagoshima University, Graduate School of Science and Engineering, Kagoshima 890-0065, Japan}
\email{yutaninm@gmail.com}

\author[0000-0002-3531-7863]{Yoshiki Toba}
\affiliation{National Astronomical Observatory of Japan, 2-21-1 Osawa, Mitaka, Tokyo 181-8588, Japan}
\affiliation{Department of Astronomy, Kyoto University, Kitashirakawa-Oiwake-cho, Sakyo-ku, Kyoto 606-8502, Japan}
\affiliation{Academia Sinica Institute of Astronomy and Astrophysics, 11F of Astronomy-Mathematics Building, AS/NTU, No.1,\\ Section 4,Roosevelt Road, Taipei 10617, Taiwan}
\affiliation{Research Center for Space and Cosmic Evolution, Ehime University, 2-5 Bunkyo-cho, Matsuyama, Ehime 790-8577, Japan}

\author[0000-0002-8779-8486]{Keiichi Wada}
\affiliation{Kagoshima University, Graduate School of Science and Engineering, Kagoshima 890-0065, Japan}
\affiliation{Research Center for Space and Cosmic Evolution, Ehime University, 2-5 Bunkyo-cho, Matsuyama, Ehime 790-8577, Japan}
\affiliation{Hokkaido University, Faculty of Science, Sapporo 060-0810, Japan}

\begin{abstract}
	\Add{In this study, we examine photoionization outflows during the late stages of galaxy mergers, with a specific focus on the relation between observed velocity of outflowing gas and the apparent effects of dust extinction.
	We used the N-body/smoothed particle hydrodynamics (SPH) code ASURA for galaxy merger simulations. 
	These simulations concentrated on identical galaxy mergers featuring supermassive black holes (SMBHs) of 10$^8$ M$_\odot$ and gas fractions of 30\% and 10\%.
	From the simulation data, we derived velocity and velocity dispersion diagrams for the AGN-driven ionized outflowing gas.
	Our findings show that high-velocity outflows with velocity dispersions of 500 km s$^{-1}$ or greater can be observed in the late stages of galactic mergers.
	Particularly, in buried AGNs, both the luminosity-weighted outflow velocity and velocity dispersion increase owing to the apparent effects of dust extinction.
	Owing to these effects, the velocity--velocity dispersion diagrams display a noticeable blue-shifted tilt in models with higher gas fractions.
	Crucially, this tilt is not influenced by the AGN luminosity but emerges from the observational impacts of dust extinction.
	Our results imply that the observed high-velocity \OIII outflow exceeding 1000 km s$^{-1}$ in buried AGNs may be linked to the dust extinction that occurs during the late stages of gas-rich galaxy mergers.}
\end{abstract}



\section{Introduction} \label{sec:intro}
Galaxy mergers are key phenomena \Add{in} understanding the evolution of galaxies and supermassive black holes (SMBHs).
Observational studies have suggested that galaxy mergers are related to the activity of galaxy centers \citep{sanders1996,farrah2002,fan2016, gao2020}.
Theoretically, galaxy mergers can enhance BH accretion and trigger active galactic nuclei (AGNs), which are often buried in a large amount of gas and dust \Add{during} the late stages of galaxy mergers \citep[e.g.,][]{narayanan2010,blecha2018,kawaguchi2020,yutani2022}.
Indeed, \citet{fan2016} reported a high merger fraction ($\sim$62\%) in heavily obscured quasars selected \Add{using a} wide-field infrared survey explorer (WISE).
In the late stages of galaxy mergers, as AGN \Add{becomes} more active, AGN feedback becomes stronger.
AGN feedback blows away the dust and gas around the AGNs, causing the buried AGNs to evolve into quasars in the late stages of galaxy mergers \citep{hopkins2008,dey2009}.
This feedback process is characterized by \Add{an} ionized gas outflow.
Thus, the properties of ionized outflows \Add{during the} late stages of galaxy mergers are crucial for understanding the formation of SMBHs and quasars.
\par

\Add{Weak} X-ray quasars ($a_{\rm ox} \leq$ -1.7) \Add{have been suggested to be} associated with strong highly ionized outflows with high equivalent widths \citep{laor2002,veilleux2022}.
\citet{toba2017} studied the ionized gas properties of 36 objects with extreme optical/infrared color (i -- [22])$_{\rm AB} >$ 7.0  and infrared bright $F_{22\mu{\rm m}}$ $>$ 3.8 mJy sources (i.e., IR-bright dust-obscured galaxies), selected by the Sloan digital sky survey and WISE.
They showed that most IR-bright dust-obscured galaxies exhibited larger velocity \Add{offsets} and larger velocity-\Add{dispersions} of \OIII than those observed in \Add{the Type-2} Seyfert galaxies.
\Add{The ionized} outflow velocities of some IR-bright DOGs exceeded 1000 km s$^{-1}$.
Theoretical models and observations suggest that IR-bright galaxies are often in the late stages of galaxy merging \citep{narayanan2010,yamada2021,yutani2022}. \par

\cite{bae2016} studied the velocity and velocity dispersion diagram (hereafter VVD diagram) of typical AGN using the three-dimensional biconical outflow model with uniform density.
In their models, dust extinction by razor\Add{--}thin disks was considered.
They showed that dust extinction strongly affected the ionized outflow velocity integrated into the line of sight.
They found that \Add{the} extinction by \Add{the} dust plane \Add{increased} the line-of-sight\Add{--}integrated velocity offset \Add{because the redshift component of biconical outflow was affected by dust extinction. This is an apparent effect\Add{,} and does not indicate a change in intrinsic outflow velocity.}
The apparent effect \Add{of} dust extinction is \Add{more} important for buried AGNs during the late stages of galaxy mergers. \par

\Add{In addition, \cite{woo2017} suggested that the bolometric luminosity of AGNs and velocity of the ionized outflow was correlated.
The relation between AGN luminosity and outflow velocity is controversial owing to the relatively small number of statistical samples; however, if there is a positive correlation between the two quantities in the late stages of the galaxy merger, stronger AGN-feedback may increase the intrinsic velocity of ionized outflows.} \par

\Add{Considering the relation between ionized outflow velocity and the apparent effects of dust attenuation or AGN activity, strong ionized outflows may accompanying AGNs in the late stages of galaxy mergers is plausible.} 
However, AGNs buried in the later stages of galaxy mergers are expected to have higher scale heights of \Add{the} dust torus and more complex outflow density distributions than \Add{the} typical AGNs considered in \cite{bae2016}.
To understand the ionized outflow associated with buried AGNs, \Add{conducting} galaxy merger simulations with sufficiently fine spatial resolution based on 3D numerical fluid dynamics simulations \Add{is necessary}.
In this study, we use the N-body/smoothed particle hydrodynamic (N-body/SPH) simulation code \texttt{ASURA} \citep{saitoh2008,saitoh2009,saitoh2013} implementing isotropic thermal AGN-feedback. \par

The remainder of this paper is organized as follows\Add{:}
\Add{In} Section \ref{sec:galaxy_merger}\Add{, we} describe the galaxy merger simulation models and \Add{the} corresponding results.
In Section \ref{sec:vvd_intro}, we summarize the VVD diagram analysis method and corresponding results.
In Section \ref{sec:discuss}, we discuss the high-velocity ionized outflow associated with \Add{the} buried AGNs identified in the observations based on our simulation results.
Finally, \Add{in} Section \ref{sec:summary} summarize the results of this study. \par
\begin{deluxetable*}{cccccccccccc}[]
	\tablenum{1}
	\tablecaption{Initial parameters of the pre-merger system\label{tab:model_parameter}}
	\tablewidth{0pt}
	\tablehead{
		Name$^{\rm a}$&${M_{BH}}^{\rm b}$&${M_{bul}}^{\rm c}$& ${M_{\rm gas}}^{\rm d}$&$\epsilon^{\rm e}$&${r_{acc}}^{\rm f}$&${r_{h}}^{\rm g}$&${R_{\rm gas}}^{\rm h}$&${\Delta M_{bul}}^{\rm i}$&${\Delta M_{\rm gas}}^{\rm j}$&${\mu_{gas}}^{\rm k}$\\
		&[$10^8 M_{\odot}$]&[$10^{11} M_{\odot}$]&[$10^{10} M_{\odot}$]&[pc]&[pc]&[kpc]&[kpc]&[$10^4 M_{\odot}$]&[$10^3 M_{\odot}$]&
	}
	\startdata
	G1 & 1.0 &  1.0 & 0.26 & 3.0 & 6.0 & 2.2 & 1.0 & 2.5 & 0.65 & 10\%\\
	G3 & 1.0 &  1.0 & 1.0 & 3.0 & 6.0 & 2.2 & 1.0 & 2.5 & 2.5 & 30\%
	\enddata
	\tablecomments{
		(a) Name of the pre-merger system.
		(b) Mass of the sink, (c) star, and (d) gas particles, respectively.
		(e) Gravitational softening.
		(f) Accretion radius of sink particle.
		(g) Effective radius \Add{of} stellar bulge Sersic profile.
		(h) Outer edge radius of uniform-density gas disk.
		(i) Mass resolution of star particles, (j) and gas particles.
		(k) Gas fraction \Add{below} 1 kpc.}
\end{deluxetable*}

\begin{figure*}[htp]
	\centering
	\includegraphics[width=\linewidth,bb=0 0 935 451]{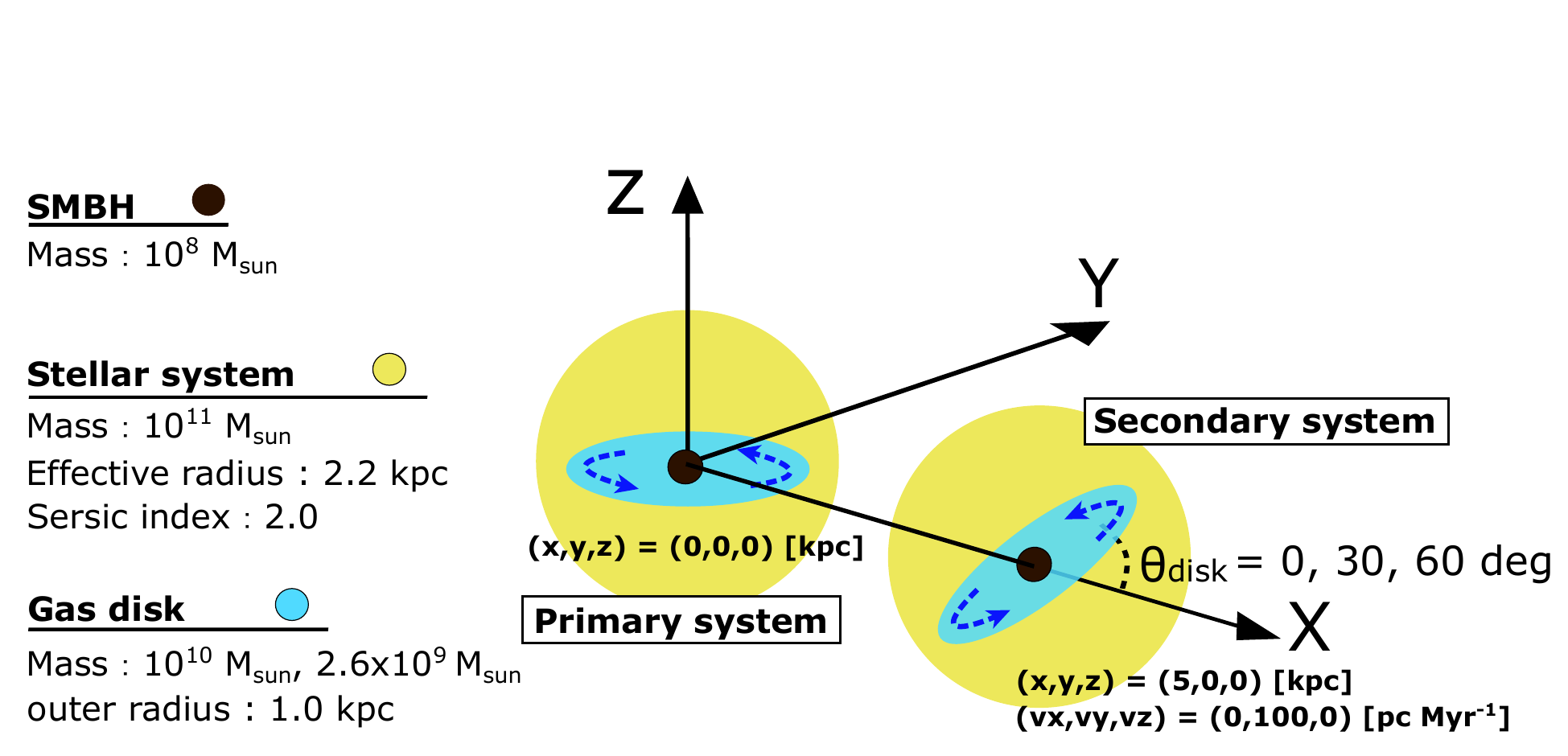}
	\caption{The schematic of the merger model\label{fig:model}}
\end{figure*}

\section{galaxy merger simulations} \label{sec:galaxy_merger}
We \Add{performed} galaxy\Add{--}merger simulations using \Add{the} N-body/SPH code \texttt{ASURA} \citep{saitoh2008,saitoh2009}.
The dynamics of the interstellar medium in \Add{a} merger system \Add{were} calculated using \textit{\Add{density-independent fomulation of smoothed particle hydrodynamics}} \citep{saitoh2013}.
\Add{The gravity} fields \Add{were} calculated using \Add{the} tree algorithm.
The tolerance parameter \Add{was} set to 0.5. \par

In this study, star formation and supernova explosions \Add{were} implemented according to \cite{saitoh2008}.
The star formation rate (SFR) is determined as $C_* \rho_{gas}/t_{ff}$, where $C_*$ is a dimensionless star\Add{--}formation efficiency parameter set to 0.0099, $\rho_{gas}$ is the local gas density, and $t_{ff}$ is the free\Add{--}fall time.
\Add{SPH particles that satisfy all four of the following conditions are converted to star particles in a stochastic manner.}
(1) \Add{The hydrogen} number density $n_{\rm H}\,>\,1000\rm{\,cm^{-3}}$\Add{.}
(2) \Add{The gas} temperature $T_{gas}\,<\,100\rm{\,K}$\Add{.}
(3) $\nabla\cdot {\mathbf{v}}_{SPH}\,<\,0$, where ${\mathbf{v}}_{SPH}$ denotes the velocity of \Add{an} SPH particle\Add{.}
(4) $\Delta Q\,<\,0$, where $\Delta Q$ denotes the thermal energy received by an SPH particle in a time step.
The initial mass function of the stellar \Add{particles} is set to \citet{salpeter1955}. 
\Add{A} probabilistic Type II supernova explosion \Add{was} introduced based on \citet{okamoto2008}.
\Add{Optically thin radiative cooling with solar metallicity is assumed for a} gas at $10\rm{\,K}\,$--$\,10^8\rm{\,K}$ \citep{wada2009}.
Far-ultraviolet radiation and photoelectric heating \Add{have} also \Add{been} considered. \par

SMBHs can \Add{acquire} the mass of gas particles within an accretion radius $r_{acc}$ \Add{satisfying the following} two conditions:
1) The kinetic energy of SPH \Add{particles} is smaller than the gravitational energy 
2) The angular momentum of \Add{an} SPH particle is \Add{less} than $J_{acc}\,=\,r_{acc}\sqrt{GM_{BH}/(r_{acc}^2\,+\,\epsilon^2)^{1/2}}$, where $\epsilon$ is the gravitational softening length.
We assume $\epsilon$ = 3.0 pc and $r_{acc}$ = 6.0 pc. \par

The isotropic AGN-feedback \Add{was} implemented by providing thermal energy ($\Delta E$ = $\eta_{AGN}\dot{M}_{BH}c^2$) weighted by the spline function to the gas particles surrounding the sink particle.
$\dot{M}_{BH}$ is the gas mass accretion rate of a BH particle per step at $r_{acc}$ and $\eta_{AGN}$ is a free parameter representing the energy-loading efficiency.
In this study, $\eta_{AGN}$ = 0.2\% \Add{was} assumed based on the discussions reported in \cite{kawaguchi2020}.

\subsection{Merger models}\label{sec:merger_model}
We \Add{investigated} the effect of a galaxy merger on the ionized gas outflow strength at the AGN origin.
We \Add{performed} a galaxy\Add{--}merger simulation focusing on the galactic nucleus at a kpc scale in the late \Add{stages} of galaxy mergers.
\Add{We assume that the bulge-core ($\sim$kpc) scale structure is not considerably distorted in the early stages of galaxy mergers.}
The parameters of \Add{pre-merger system} are reported in Table \ref{tab:model_parameter}.
\Add{The pre-merger system comprises a supermassive BH, stellar bulge, and gas disk with a kpc-scale radius.
We are interested in the high-velocity outflow observed by \cite{toba2017}.
The SMBH masses in their samples were distributed at approximately 10$^8M_\odot$; therefore, we set the SMBH mass in our models to 10$^8M_\odot.$
The stellar bulge mass was determined using the Magorian relation.
Because we focus on gas-rich systems, the mass profile of the stellar bulge is based on star-forming galaxies with a redshift of $\sim$ 2.
According to observational studies of star-forming galaxies, we have set the effective radius at 2.2 kpc and the Sersic index at 2.0 for the stellar bulge \citep{barro2017, Paulino-Afonso2017}.
The distribution function for the stellar bulge was obtained from \texttt{AGAMA}, which provides a self\Add{--}consistent N-body system \citep{vasiliev2019}.}
The total gas mass is set \Add{such} that the gas \Add{fractions} $\mu_{gas}=M_{gas}/(M_{gas}+M_{bul}(r_{xy} < 1\,{\rm kpc}))$ \Add{are} 10\% and 30\%, \Add{respectively}, where $r_{xy}$ is the radial length in cylindrical coordinates with its origin at the SMBH.
$M_{bul}(r_{xy} < 1\,{\rm kpc})$ is the \Add{bulge} mass \Add{within} the bulge \Add{of} a \Add{1--}kpc radius.
The gas fraction is based on observations of star forming galaxies \cite{erb2006}.\par

We \Add{simulated} collisions between identical systems.
As the schematic in Figure \ref{fig:model} shows, the systems \Add{were} separated from each other by 5 kpc and given relative velocities of 100 pc/Myr in the y-direction.
\Add{We are interested in the late stages of galaxy mergers, where the two galactic nuclei share a common envelope \citep{stierwalt2013}. 
Given that the effective radius of the bulge was 2.2 kpc and the initial radius of the disk was 1 kpc, we separated it by 5 kpc to prevent contact and initial mass accretion to the nucleus.}
In a collisional system, the initial spin angular momentum vector of the gas disk satisfies 
\begin{equation}
	L_z^{prime} = L_z^{second}\cos{\theta_{disk}}-L_x^{second}\sin{\theta_{disk}},
\end{equation}
where $\theta_{disk}$ \Add{denotes} the angle between the gas disk of the secondary system and xy-plane.
We \Add{considered} three cases with $\theta_{disk}$ = 0 \Add{deg}, 30 \Add{deg}, and 60 deg.
In the following discussion, G3T0 denotes the case of $\theta_{disk}=0$ deg for G3 with a gas fraction of 30\% and coalescence between identical systems.
We \Add{simulated} six models: G1T0, G1T30, G1T60, G3T0, G3T30, and G3T60 as \Add{listed} in Table \ref{tab:sixmodels}.

\begin{deluxetable}{ccccccccc}[]
	\tablenum{2}
	\tablecaption{Free parameters of the six models\label{tab:sixmodels}}
	\tablewidth{0pt}
	\tablehead{&Model name &&&${\mu_{gas}}^{\rm a}$&&&${\theta_{disk}}^{\rm b}$&}
	\startdata
	& G1T0 &&& 10\% &&& 0 deg &\\
	& G1T30 &&& 10\% &&& 30 deg &\\
	& G1T60 &&& 10\% &&& 60 deg &\\
	& G3T0 &&& 30\% &&& 0 deg &\\
	& G3T30 &&& 30\% &&& 30 deg &\\
	& G3T60 &&& 30\% &&& 60 deg &
	\enddata
	\tablecomments{
		(a) \Add{The gas} fraction within 1 kpc, i.e., $\mu_{gas} = M_{gas}/(M_{gas}+M_{bul}(r_{xy} < 1\,{\rm kpc}))$.
		(b) Angle between gas disk of secondary system and xy-plane.}
\end{deluxetable}

\begin{figure*}[h]
	\centering
	\includegraphics[width=0.8\linewidth,bb = 0 0 2770 3687]{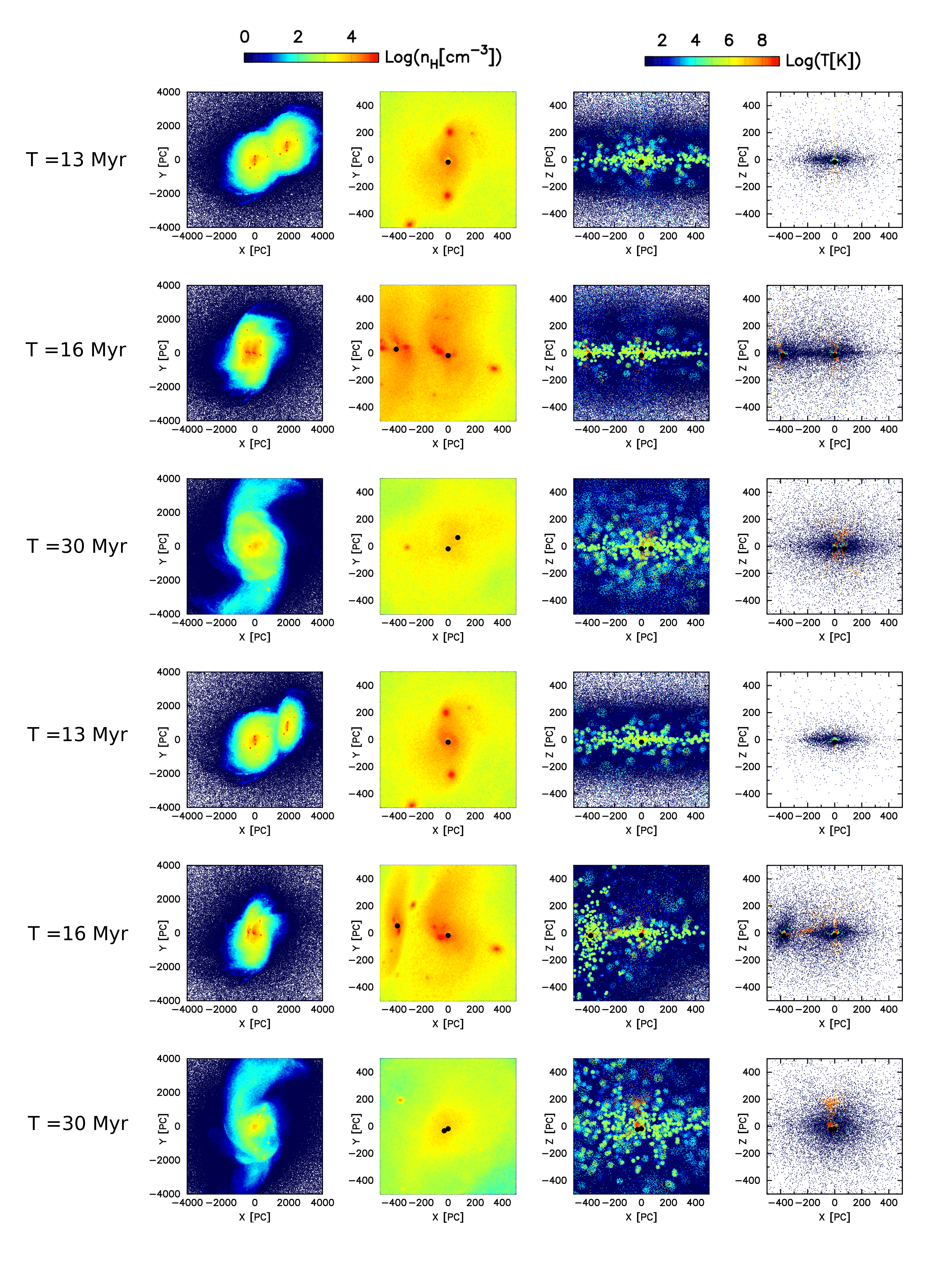}
	\caption{Map of gas density (1st and 2nd columns from the left) and gas temperature (3rd and 4th columns from the left). 
	The rightmost column shows only gas particles that received AGN-feedback energy within a Myr.
	The top three rows show G3T0, while the bottom three rows show G3T60. 
	\label{fig:density}}
\end{figure*}

\subsection{\Add{Galaxy} merger simulation results}\label{sec:Dbh}
Figure \ref{fig:density} shows the evolution of the gas density and temperature distribution for G3T0 and G3T60.
As the system approaches, the gas is compressed and sink particles are buried in the dense gas region. 
The mass accretion rate to the sink particle is enhanced in this stage. 
In addition, the \Add{two} right columns in Figure 2 show the temperature distribution of the gas, and the rightmost column shows the temperature distribution of the gas particles that received AGN-feedback energy within Myr.
The temperature distributions indicate that \Add{the} hot gas particles are blown vertically up from the disk by the AGN-feedback. \par

\begin{figure}[h]
	\centering
	\includegraphics[width=8cm,bb = 0 0 720 936]{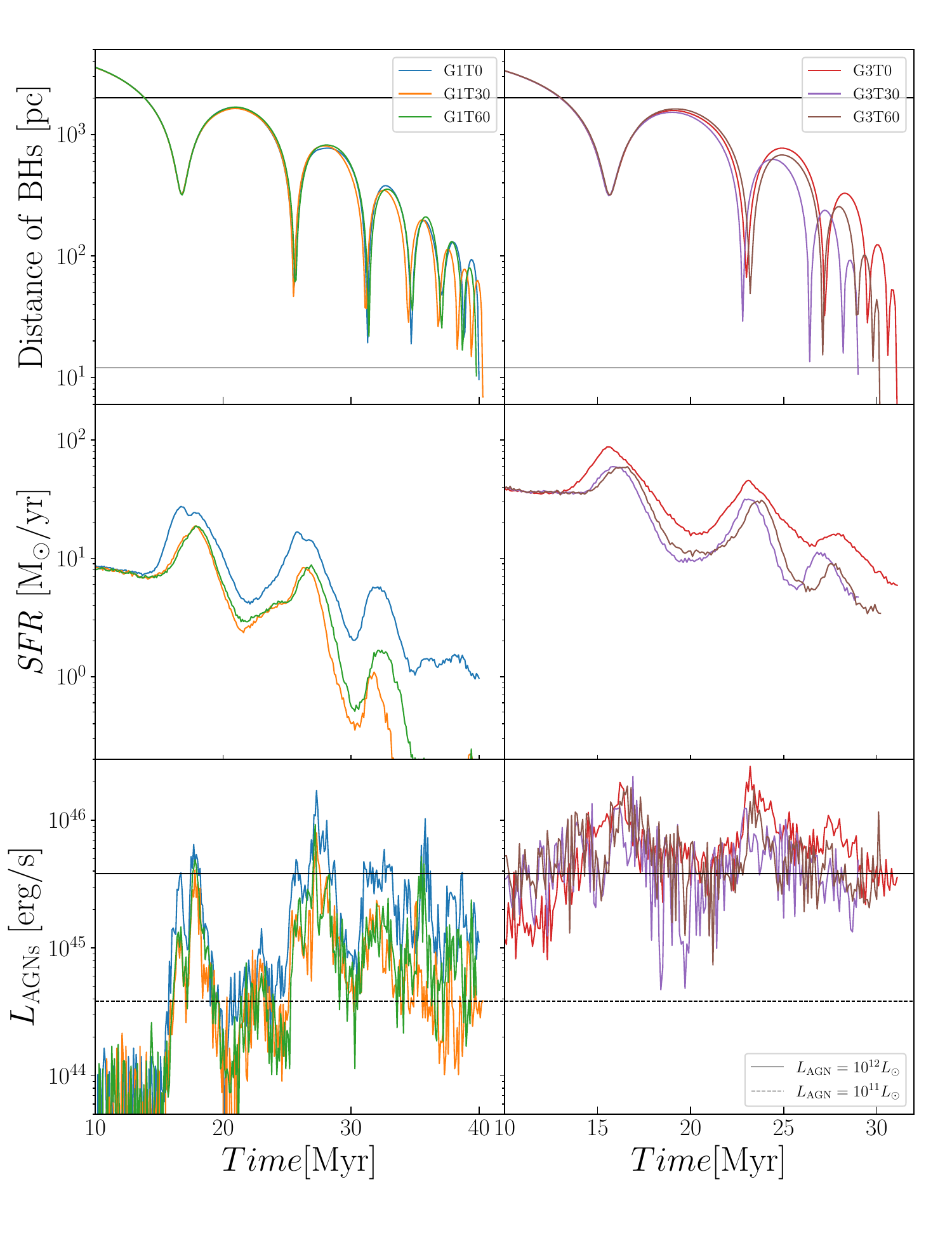}
	\caption{The first row shows the distance between binary BHs. \Add{The second row} denotes the star formation rage (SFR). \Add{The third row} shows the total bolometric luminosity of the two AGNs ($L_{AGNs} = \Sigma\, 0.1*\Dot{M}_{sink}c^2$). \Add{In the first to third rows, the} left column is for G1T0, G1T30, and G1T60; the right column is for G3T0, G3T30, and G3T60. \label{fig:Dbh}}
\end{figure}

Figure \ref{fig:Dbh} shows the evolution of SMBH binary distance, total bolometric AGN-luminosity, and star formation rate for \Add{the} three models.
\Add{The} top row of Figure \ref{fig:Dbh} \Add{shows} that the SMBH binary orbits are similar in all three models with the same gas fraction.
The middle row in Figure \ref{fig:Dbh} shows that there \Add{was} no significant difference in AGN luminosity among the three models with the same gas fraction (models G1 and G3).
In all three models, the AGN activity and star formation activity \Add{were} enhanced around the first and second pericenters of the BH-binary orbit.
AGN luminosity and star formation rate \Add{tended} to be higher with a gas fraction of 30\% than with 10\%.
For a gas fraction \Add{of} 30\%, the AGN bolometric luminosity \Add{ranges from} $10^{11} L_{\odot}$ to $10^{13} L_{\odot}$, \Add{whereas} for a gas fraction 10\%, the AGN luminosity \Add{ranges from} $10^{10} L_{\odot}$ to $10^{12}L_{\odot}$.

\section{Velocity and velocity dispersion diagram}\label{sec:vvd_intro}
\subsection{Integral of photoionized outflow velocity}
We \Add{used} the \textit{velocity and velocity-dispersion (VVD) diagrams} to investigate the relation between \Add{the} galaxy merger and ionized outflow velocity.
VVD diagrams \Add{have been} used to investigate the statistical properties of \Add{the} ionized outflow velocity in AGNs \citep{woo2017,rakshit2018}.
We \Add{constructed} the VVD diagrams using \Add{the} intrinsic velocities of SPH particles derived from the simulated galaxy mergers data to avoid the high computational cost of 3D ionizing emission line pseudo-observations. \par

In a 3D polar coordinate system centered on each sink particle, we \Add{selected} photoionized gas particles from those above 8000 K by \Add{limiting} them \Add{to} typical \OIII ionization parameters (-3 $<$ $\log$U $<$ -1)\Add{.}
\begin{equation}
	U = \frac{\int^{\infty}_{\nu_0}L_\nu e^{-\tau_\nu}/h\nu d\nu}{4\pi r^2 cN_e},\label{eq:U}
\end{equation}
where $N_e$ is the number density of SPH particles \Add{at} temperatures above 8000 K.
The ionization parameter U is defined for an SPH particle as the ratio of the number density of ionizing photons with energy $>$ 13.6 eV to \Add{the} electron number density $N_e$ \citep[e.g.,][]{wada2018}.
We consider the radiation emitted from \Add{the} AGN as \Add{the} central point source.
The AGN spectral energy distribution (SED) ($f_\nu = L_\nu/4\pi r^2$) was obtained from Cloudy's AGN table \citep{ferland2017}.
The AGN SED \Add{was} assumed \Add{to be}
\begin{equation}
	f_\nu = \nu^{\alpha_{uv}}\exp(h\nu/kT_{BB})\exp(-kT_{IR}/h\nu)+b\nu^{\alpha_x},
\end{equation}
Here, $\alpha_{uv} = -0.5$, $T_{BB}=1.5\times10^5$ K, $kT_{IR} = 0.01$ Ryd, $\alpha_x$ = -1.0, and b is a constant yielding $\alpha_{ox}$ = -1.4.
The AGN bolometric luminosity depends on the mass accretion rate on \Add{the} sink particle.
For \Add{the} optical depth ($\tau_{\nu}$) expressed in \Add{Equation} \ref{eq:U}, we consider scattering and absorption by dust, photoionization of hydrogen, and Thomson scattering:
\begin{align}
	\tau_{\nu} = \sigma_{\nu,dust}N_{{\rm H},\Add{agn}} + &\sigma_{\nu,H}N_{{\rm H},\Add{agn}}(<8000{\rm K})\notag\\
	&+\sigma_{T}N_{{\rm H},\Add{agn}}(\geq8000{\rm K})
\end{align}
where $\sigma_{\rm dust}$ is \Add{the} extinction cross section of MRN dust \citep{laor1993}, $\sigma_{\rm H}$ is \Add{the} photoionization cross\Add{--}section of hydrogen \citep{osterbrock2006}, and $\sigma_{\rm T}$ is $6.65\times10^{25}$cm$^2$, which is \Add{the} Thomson scattering cross\Add{--}section.
\NHagn is the column density from the SMBH to the SPH particle.
In addition, we \Add{extracted} gas particles \Add{not bound} to the SMBH potential.
Because we \Add{were} interested in the outflow component, we \Add{used the} radial velocity in a 3D polar coordinate system centered on each sink particle. \par

From the gas particles selected \Add{using} the U parameter (-3 $< \log{U} <$ -1), we \Add{selected} gas particles that \Add{were} not trapped by an SMBH potential and \Add{calculated} the line-of-sight velocity and velocity dispersion.
\Add{To assess whether gas particles are trapped by SMBH potential, the gas particles velocities are calculated by $\sqrt{v^2_r + C_s^2}$, where $v_r$ is radial velocity from SMBH and $C_s$ is sound speed. If the velocity of a gas particle is exceeds $\sqrt{2GM_{BH}/r}$, then its velocity contributes to Equations \ref{eq:voff} and \ref{eq:sigv}.}
We assume that the \OIII line intensity is proportional to ${n_H}^2$ because the \OIII line is the collision\Add{--}excited line.
In \Add{the} cartesian coordinate system centered on each sink particle, we calculate mean velocity and velocity dispersion as 
\begin{align}
	\overline{v}_{los}\Add{(\theta,\phi)} &= \frac{\Sigma_i {n_H}_i^2 e^{-{\tau_{5007}}_i}{v_{los}}_i\Add{(\theta,\phi)}}{\Sigma_i {n_{H}}_i^2 e^{-{\tau_{5007}}_i}}\label{eq:voff}\\
	\overline{\sigma}^2_{v_{los}}\Add{(\theta,\phi)} &= \frac{\Sigma_i {n_H}_i^2 e^{-{\tau_{5007}}_i}{v_{los}}_i^2\Add{(\theta,\phi)}}{\Sigma_i {n_H}_i^2 e^{-{\tau_{5007}}_i}}-\overline{v}^2_{los}\Add{(\theta,\phi)}\label{eq:sigv},
\end{align}
\Add{where ${v_{los}}_i$ is the line-of-sight velocity of each SPH particle in a rest frame centered on sink particle and $\tau_{5007}$ equals to $\sigma_{dust} (\lambda=5007{\rm\AA}){N_{{\rm H},los}}(\theta_0,\phi_0)$.
${N_{{\rm H},los}}(\theta_0,\phi_0)$ is the column density from each SPH particle to r = 6 kpc in the ($\theta_0,\phi_0$) direction.
The $\theta_0$ is the angle from the rotation axis of the primary system (i.e., z-axis in Figure \ref{fig:model}).
We use ${N_{{\rm H},los}}(\theta_0,\phi_0)$ as the approximate column density only for the line of sight within the solid angle $\Omega = 2\pi(1-\cos(\pi/12))$ sr from ($\theta_0$,$\phi_0$).} \par

\Add{We have fixed $\phi_0$ = 0 deg (i.e., in the x-axis direction in Figure \ref{fig:model}) and calculated for four cases where $\theta_0$ is 0 deg, 30 deg, 60 deg, and 90 deg (i.e., covering the range from $\theta$ = -15 deg to 105 deg and $\phi$ -15 deg to 15 deg).
In our simulations, the OIII ionization outflow of the AGN origin was on the scale of a few hundred pc or less (see Figure \ref{fig:hist}), and the gas density distribution in the $\phi$ direction did not change significantly on that scale (see Figure \ref{fig:density}).}
\Add{We choose 360 lines of sight within $\Omega = 2\pi(1-\cos(\pi/12))$ for each $\theta_0$ with equal $\sin{\theta}d\theta d\phi$ in each SMBH.
As there are two SMBHs in a model, we plotted $2\times4\times$360 points per snapshot on the VVD diagram for all models.}\par

To assess the relation between galaxy mergers and ionized outflows \Add{accurately}, we define \textit{the merger phase} based on the distance between binary BHs.
We are interested in the stage \Add{in which} the gas disks merge with each other.
Because the initial gas disk radius is 1 kpc and BH accretion radius is 6 pc (see Table \ref{tab:model_parameter}), the merger phase is defined as the distance between the binary BHs between 2 kpc and 12 pc.
\Add{The reason for setting 2 kpc is that if the distance between SMBHs is less than 2 kpc, two bulge systems with an effective radius of 2.2 kpc strongly gravitationally interact and the gas disk with a radius of 1 kpc is incredibly distorted and the mass accretion rate to the SMBHs increases.}\par

\begin{figure}[h]
	\centering
	\includegraphics[width=8cm,bb = 0 0 360 720]{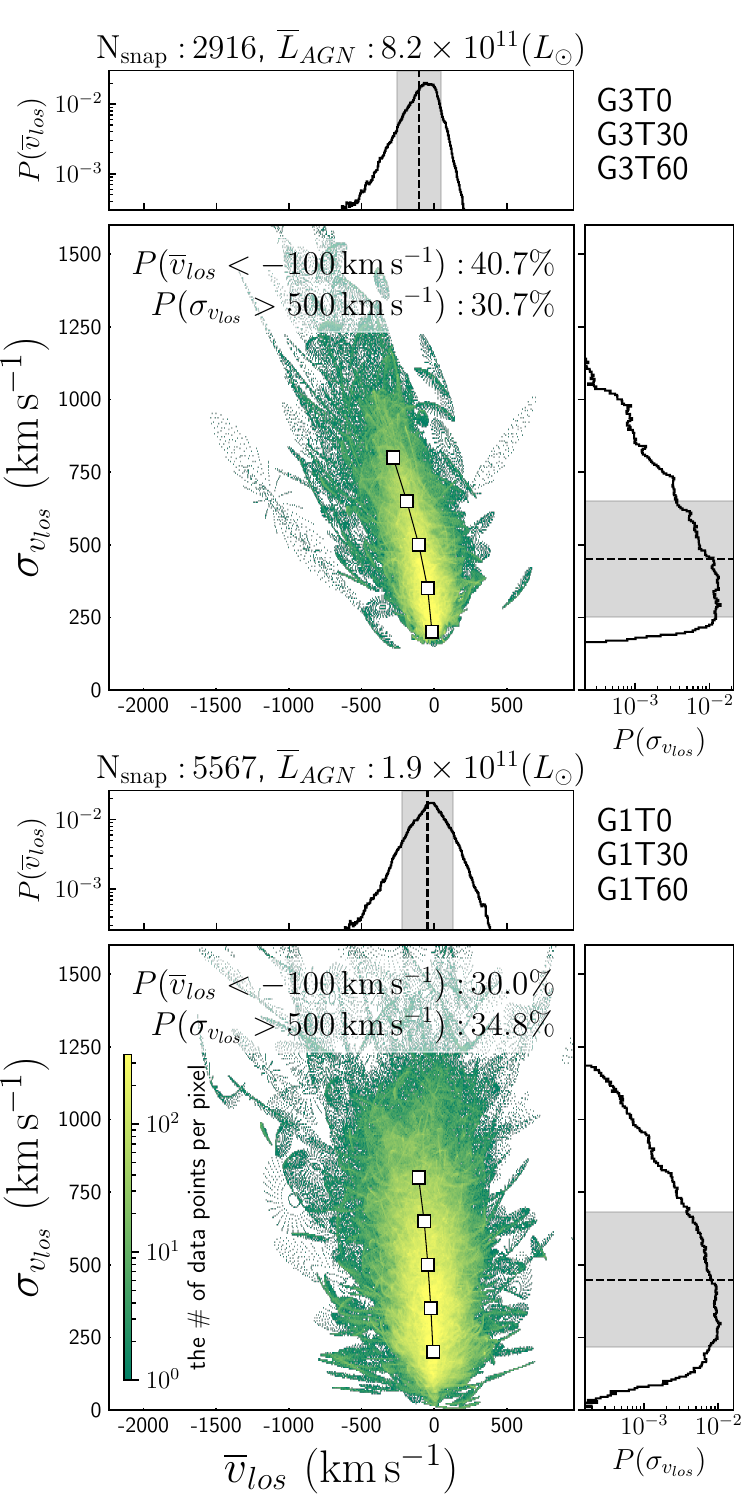}
	\caption{Comparison of VVD diagrams for gas fraction of 30\% (upper row), and gas fraction of 10\% (bottom row). \Add{The upper panel includes three models, namely G3T0, G3T30, and G3T60. The lower panel includes three models, namely G1T0, G1T30, and G1T60.} The dotted line in the histogram for each axis indicates the mean value, and the gray background color indicates the standard deviation. $N_{snap}$ is the total number of snapshots in the VVD diagram; all points in the VVD diagram are 360$\times N_{snap}$. $\overline{L}_{AGN}$ is the mean luminosity of the snapshots used in each panel. \label{fig:vvd_merger}}
\end{figure}

\begin{figure}[h]
	\centering
	\includegraphics[width=8cm,bb = 0 0 360 720]{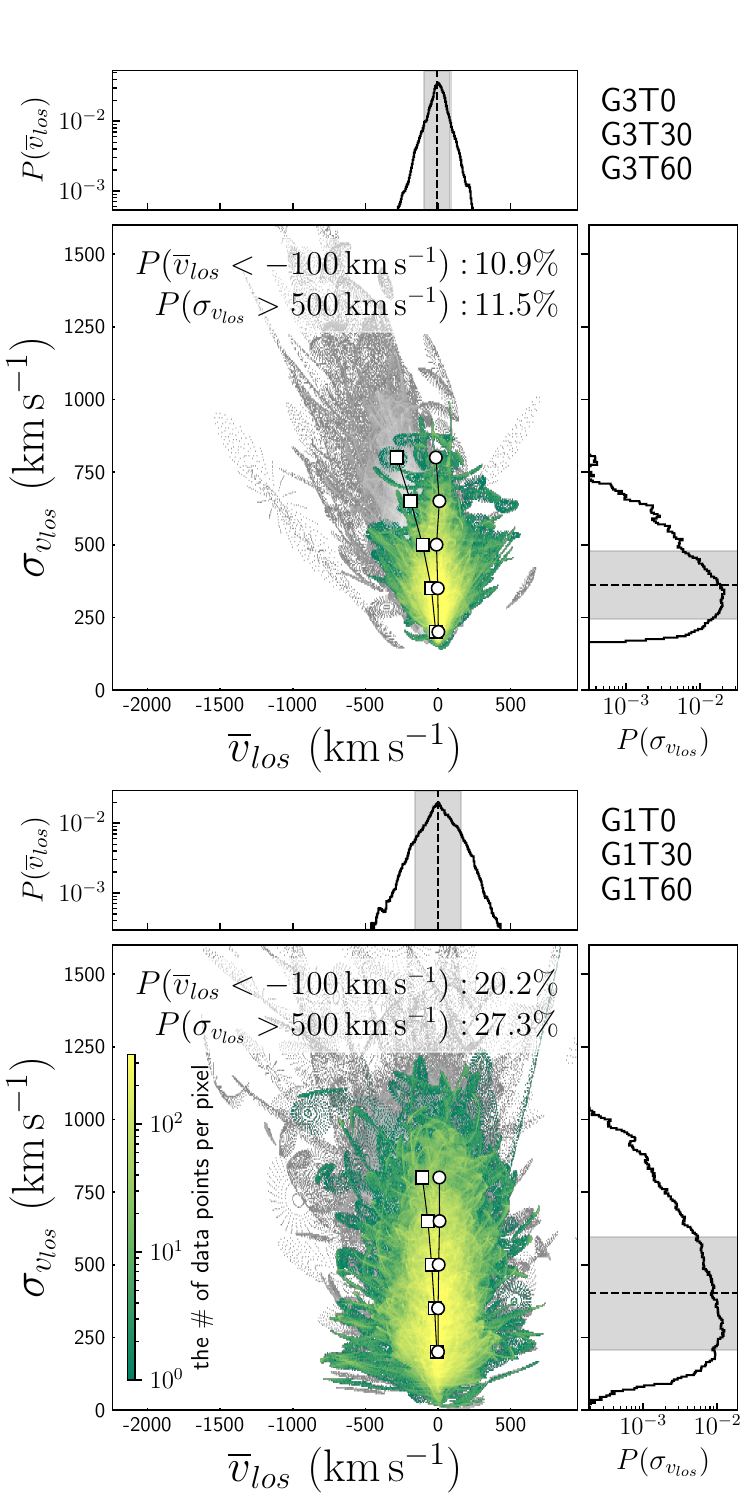}
	\caption{VVD diagrams for galaxy merger models with $\tau_{5007}$ = 0 in 
		equations \ref{eq:voff} and \ref{eq:sigv} for a gas fraction of 30\% (upper panel) and a gas fraction of 10\% (bottom panel), respectively.
		The gray plot denotes the VVD considering the dust extinction in Figure \ref{fig:vvd_merger}.
		The mean value for \voff of each \sigv bin is indicated by a circle ($\tau_{5007}$=0) and square (\Add{gray plot}).
		The dotted line in the histogram for each axis indicates the mean value, while the gray background color indicates the standard deviation for galaxy merger models with $\tau_{5007}$ = 0.
		\label{fig:vvd_tau}}
\end{figure}

We \Add{plotted} our models on \Add{a} VVD diagram for snapshots \Add{during} the merger phase.
\Add{However, the following two cases are not plotted in the VVD diagram: 1) when $L_{agn}$ = 0, and 2) when Equation \ref{eq:h} is not satisfied.
\begin{align}
	\tilde{h} = \frac{\Sigma_i {n_H}_i^2 e^{-{\tau_{5007}}_i}h_i}{\Sigma_i {n_{H}}_i^2 e^{-{\tau_{5007}}_i}} \leq 12\,{\rm pc},\label{eq:h}
\end{align}
where $h_i$ is the kernel radius of SPH particles.
As the accuracy of SPH particles depends on the local particle density, this condition eliminates snapshots that are heavily influenced by noisy SPH particles.
In our calculations, the gravity softening was 3 pc; therefore, we set the threshold at four times the gravity softening (12 pc).
For G3T0 model, the merger phase ranged from 13.0 to 30.1 Myr, and snapshots were obtained every 0.1 Myr.
Therefore, at the maximum 181 snapshots were plotted on the VVD diagram.
$2\times4\times360$ lines of sight were selected for each snapshot in all the models.
Thus, for the G3T0 model, at the maximum 181$\times$2880 points were plotted on the VVD diagram.}\par

\subsection{VVD diagrams of gas-rich galaxy merger models}\label{sec:vvd_voff}
\Add{Figure 4 shows VVD diagrams for merger models with gas fractions of 30\% and 10\%.
These VVD diagrams show an ionization outflow component with a velocity dispersion exceeding 500 km s$^{-1}$ regardless of the gas fraction, confirming the high-velocity outflow from the galaxy merger simulations.
In addition, the model with a gas fraction of 30\% is more tilted toward the blue-shifted side than that with a gas fraction of 10\%.
Indeed, the percentage of samples that were blue-shifted more than -100 km s$^{-1}$ was 40.7\% for the 30\% gas fraction and 30.0\% for the 10\% gas fraction model.
According to \cite{bae2016}, the tilt in the VVD diagram is produced by dust extinction against the redshift component of the dipole outflow.
In this study, we investigated the effect of dust extinction on the line-of-sight average velocity and velocity dispersion (equations \ref{eq:voff} and \ref{eq:sigv}) during the late stages of galaxy mergers.}

\begin{figure}[h]
	\centering
	\includegraphics[width=8cm,bb = 0 0 360 720]{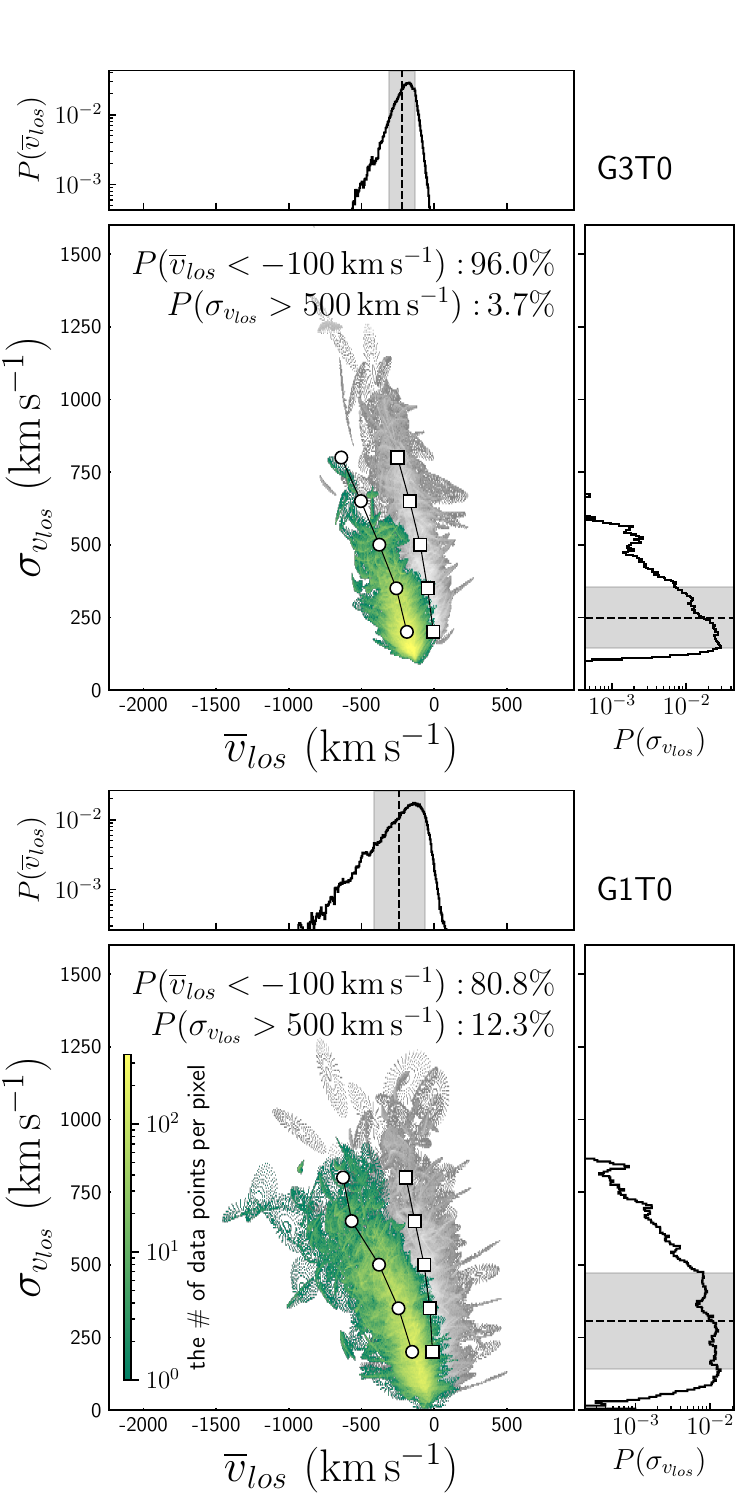}
	\caption{
		\Add{VVD diagram with $\tau_{5007}$ = 0 for $v_z$ $>$ 0 particles and $\tau_{5007}$ = $\infty$ for $v_z$ $\leq$ 0 with rest frame for SMBH for a G3T0 with a gas fraction of 30\% (upper panel) and a G1T0 with a gas fraction of 10\% (bottom panel), respectively.
			The gray plot denotes the VVD considering the dust extinction for all ionized outflow gas particles.
			The mean value for \voff of each \sigv bin is indicated by a circle for the colored VVD diagram and square for the gray VVD diagram.
			The dotted line in the histogram for each axis indicates the mean value while the gray background color indicates the standard deviation of the colored VVD diagram.}
		\label{fig:vvd_tau_blue}}
\end{figure}

\begin{figure}[h]
	\centering
	\includegraphics[width=8cm,bb = 0 0 360 720]{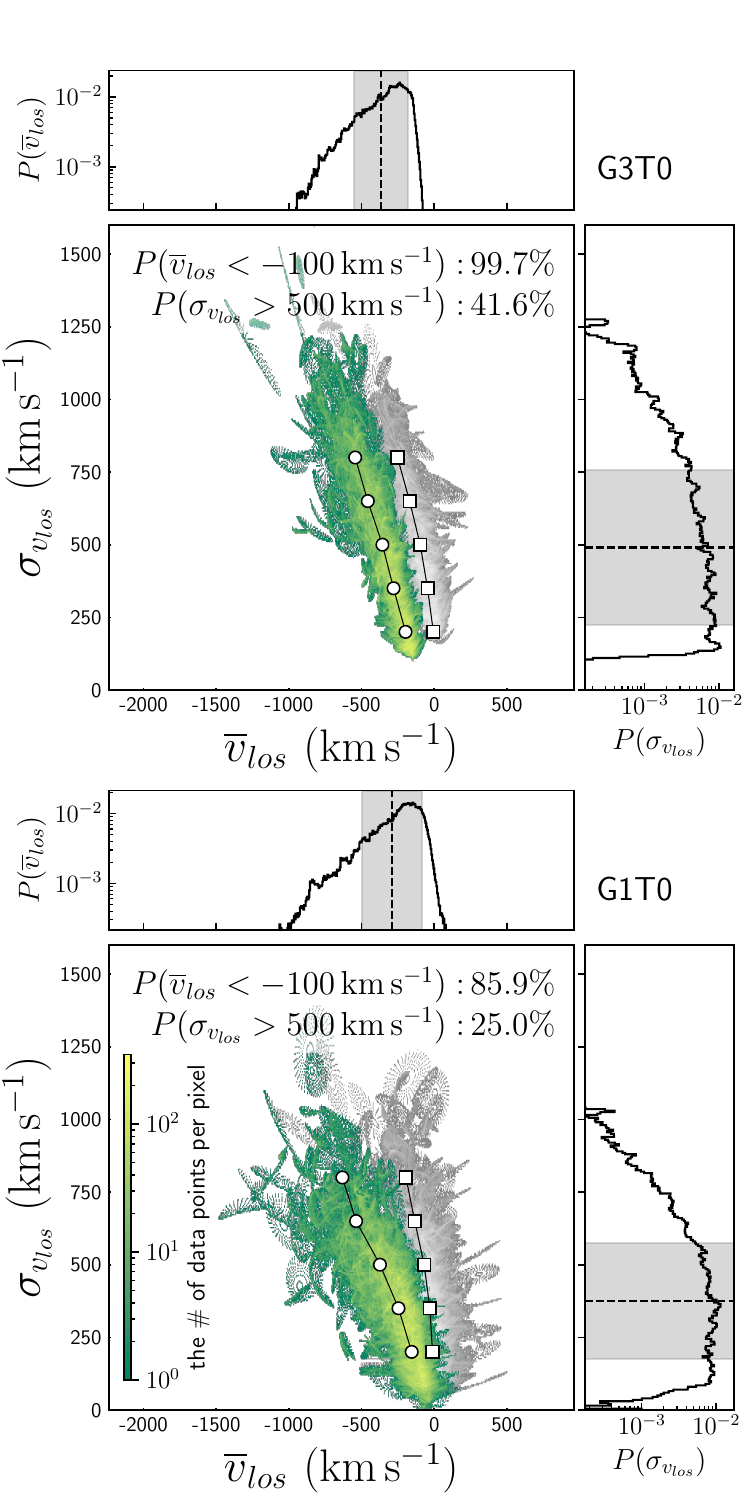}
	\caption{
		\Add{VVD diagram with $\tau_{5007}$ = $\infty$ for $v_z$ $\leq$ 0 with rest frame for SMBH for a G3T0 with a gas fraction of 30\% (upper panel) and a G1T0 with a gas fraction of 10\% (bottom panel), respectively.
		\Adds{Here dust extinction is considered for $v_z$ $>$ 0 particles.}
		The gray plot denotes the VVD considering the dust extinction for all ionized outflow gas particles.
		The mean value for \voff of each \sigv bin is indicated by a circle for the colored VVD diagram and square for the gray VVD diagram.
		The dotted line in the histogram for each axis indicates the mean value while the gray background color indicates the standard deviation for the colored VVD diagram.}
		\label{fig:vvd_tau_dence}}
\end{figure}

\subsubsection{Apparent effect of dust extinction}\label{sec:vvd_tau}
Figure \ref{fig:vvd_tau} plots the VVD diagram with $\tau_{5007}$ = 0 expressed in Equations \ref{eq:voff} and \ref{eq:sigv}.
Figure \ref{fig:vvd_tau} \Add{confirms} that a symmetric distribution with respect to \voff = 0.
\Add{Since the radiation emitted from all ionized gases is optically thin  in Figure \ref{fig:vvd_tau}, the redshift component will cancel out the blueshift component.
Figure \ref{fig:vvd_tau_blue} shows the VVD diagram with $\tau_{5007}$ = 0 for $v_z$ $>$ 0 particles and $\tau_{5007}$ = $\infty$ for $v_z$ $\leq$ 0 with rest frame for SMBH.
In Figure \ref{fig:vvd_tau_blue}, for simplicity, VVD diagrams are plotted from a face-on view (0 $\leq$ $\theta$(deg) $\leq$ 45) for coplanar merger models (G3T0 and G1T0).
Figure \ref{fig:vvd_tau_blue} shows the effect of dust extinction for the redshift component on the VVD diagram.
We confirmed that the tilt in the VVD diagram was produced by dust extinction for the redshift component.
This result is consistent with those \cite{bae2016}.}

\begin{figure}[h]
	\centering
	\includegraphics[width=8cm, bb=0 0 461 346]{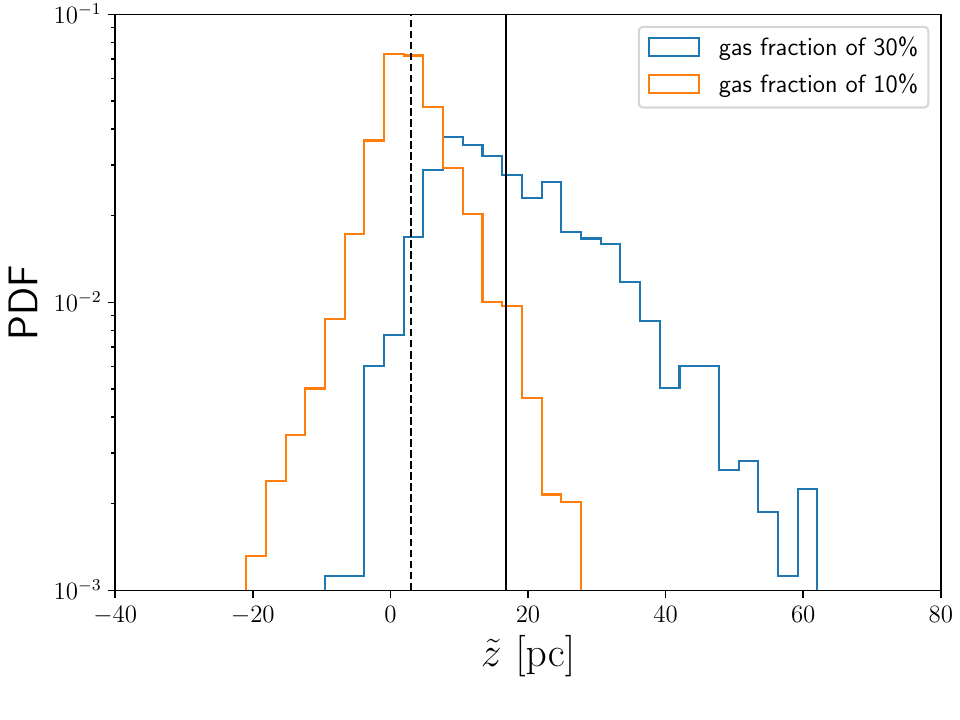}
	\caption{\Adds{Probability density function of equation \ref{eq:z} for the merger phase of models with gas fraction of 30\% which are G3T0, G3T30, and G3T60 (blue) and 10\% which are G1T0, G1T30, and G1T60 (orange) from face-on view ($\theta \leq$  45 deg). The vertical solid and dashed lines correspond to the medians of the histgrams with gas fraction of 30\% and 10\%, respectively.}\label{fig:hist}}
\end{figure}

\Add{Figure \ref{fig:hist} plots the probability density function (hereafter, PDF) of equation \ref{eq:z} \Adds{for the six models (i.e., G3T0, G3T30, G3T60, G1T0, G1T30, and G1T60)} from face-on view ($\theta\leq$ 45 deg).
\begin{align}
	\tilde{z} = \frac{\Sigma_i {n_H}_i^2 e^{-{\tau_{5007}}_i}z_i}{\Sigma_i {n_{H}}_i^2 e^{-{\tau_{5007}}_i}},\label{eq:z}
\end{align}
where dust extinction was considered (i.e., $\tau_{5007}\neq0$).
Figure \ref{fig:hist} shows that the PDF with a gas fraction of 30\% exhibits a systematic offset of z $>$ 0 (observer side).
Although for G1T0 with a gas fraction of 10\%, the PDF is symmetric for z $\simeq$ 0, and for G3T0 with a gas fraction of 30\%, it is drastically reduced for z $\lesssim$ 0.
The offset was caused by dust extinction of the redshift component.
As shown in Figure \ref{fig:vvd_tau_blue}, when the redshift component was obscured by dust, the VVD was tilted.
Therefore, when the gas fraction is 10\%, the VVD diagram is symmetrical because it is optically thin with respect to the redshift component (i.e., beyond the disk).}\par

\Add{In addition, dust extinction is important for \sigv with a gas fraction of 30\% (see Figure \ref{fig:vvd_tau}).
For the models with a gas fraction of 30\%, Figure \ref{fig:vvd_tau} shows that \pp{\sigv}{500} is \Add{0.37} times lower without dust extinction than that with dust extinction.
This was caused by the density gradient in the line of sight (i.e., the vertical density gradient).
In Figure \ref{fig:vvd_tau}, the dense gas in the optically thick region of the gas disk affects Equations \ref{eq:voff} and \ref{eq:sigv} by the square of the density.
Such high-density gas regions have lower temperatures and velocity dispersions; therefore, the velocity dispersion in Figure \ref{fig:vvd_tau} is systematically lower.}\par

\Add{Figure \ref{fig:vvd_tau_dence} shows that the VVD diagram with $\tau_{5007}$ = $\infty$ for $v_z$ $\leq$ 0 particles with rest frame for SMBH.
In contrast, for $v_z$ $>$ 0 particles, dust extinction was considered (i.e, $\tau_{5007}\neq0$, which is an important difference from Figure \ref{fig:vvd_tau_blue}).
Figure \ref{fig:vvd_tau_dence} shows a systematically higher velocity dispersion than that shown in Figures \ref{fig:vvd_tau} and \ref{fig:vvd_tau_blue}.
As high-density regions are strongly affected by dust extinction, high-temperature, high-velocity, and optically thin regions will be more significant.
Consequently, in the optically thin region, a relatively dense region was observed to be the most violent collision--excited.
Therefore, in buried AGNs, the velocity dispersion of the observed emission lines is higher, owing to dust extinction.
Note that in a typical AGN, the smaller the effect of dust extinction, the more the velocity dispersion increases, owing to the contribution of the redshift component \citep{bae2016}.
The increase in velocity dispersion due to dust extinction is not important in a typical AGN, as shown in the VVD diagram for a gas fraction of 10\% in Figure 5.}\par

\begin{figure*}[h]
	\centering
	\includegraphics[width=16cm, bb=0 0 792 792]{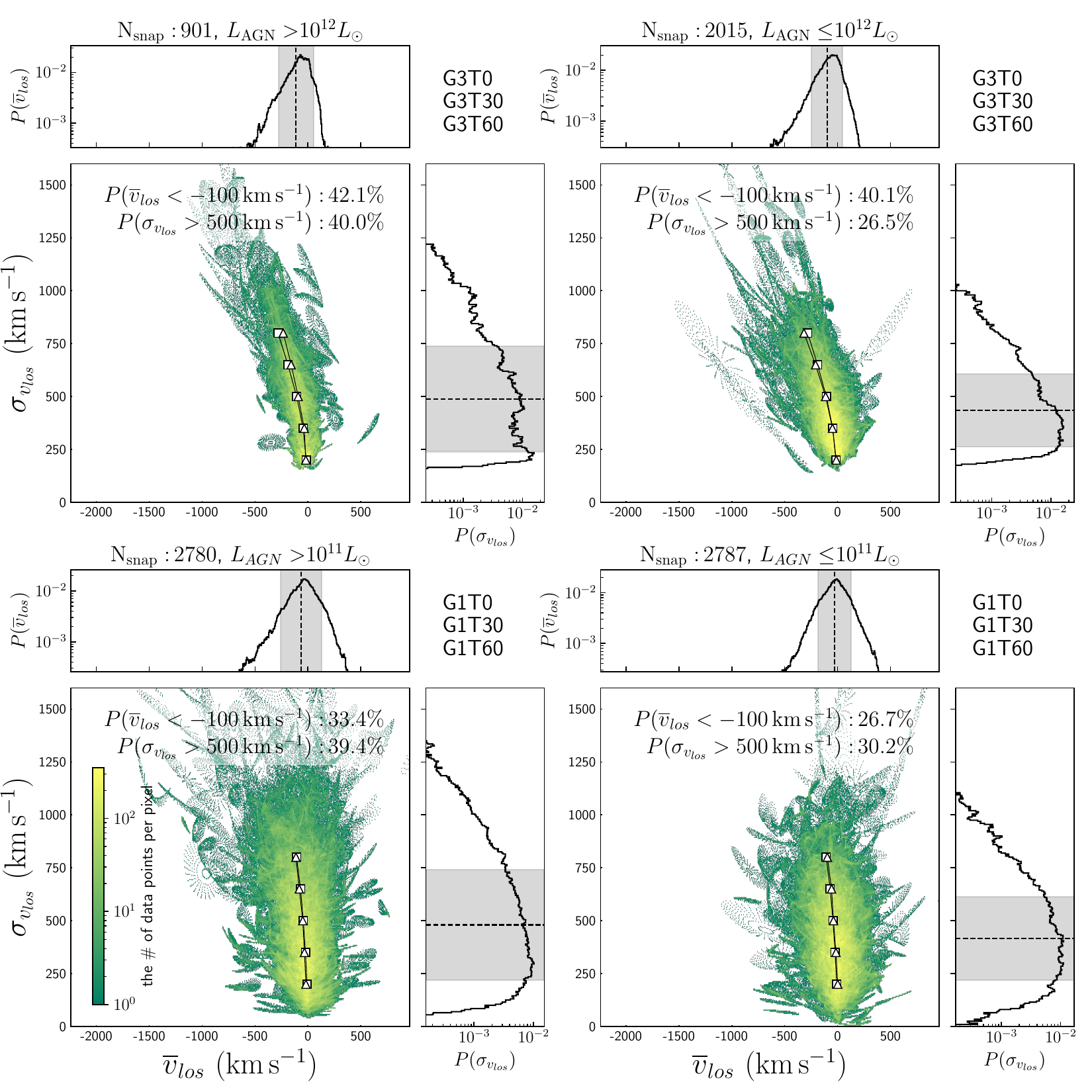}
	\caption{\Add{VVD diagrams classified with respect to AGN-luminosity for a gas 
		fraction of 30\%, namely, G3T0, G3T30, and G3T60 (upper row), and gas fraction of 10\%, namely, G1T0, G1T30, and G1T60 (bottom row).
		\Adds{The AGN luminosity of individual SMBHs was used, not the combined luminosity of each pair as reported in Figure \ref{fig:Dbh}.}
		For a gas fraction of 30\%, the threshold was set to 10$^{12}$ L$_\odot$, while for a gas fraction of 10\%, the threshold was set to 10$^{11}$ L$_\odot$.
		The mean value for \voff of each \sigv bin is indicated by square (all \Add{AGN-luminosity}, i.e., Figure \ref{fig:vvd_merger}) and triangle symbols (each AGN-luminosity bin).
		The dotted line in the histogram for each axis indicates the mean value while the gray background color indicates the standard deviation.}
		\label{fig:vvd_agn}}
\end{figure*}

\section{Discussion} \label{sec:discuss}
\subsection{AGN-luminosity and VVD diagrams}\label{sec:vvd_agn}
\Add{Figure \ref{fig:vvd_agn} shows the VVD diagrams depicted in Figure \ref{fig:vvd_merger}, classified with respect to the AGN luminosity.
The upper row in \Adds{Figure \ref{fig:vvd_agn}} shows the VVD diagram for a gas fraction of 30\%, whereas the bottom row shows the diagram for a gas fraction of 10\%.
The tilt toward the blueshift of the VVD diagram is not significantly changed against AGN luminosity, compared to the effect of dust extinction in Figure \ref{fig:vvd_tau}.
	Thus, the AGN luminosity does not significantly affect the tilt of the VVD diagram.}\par

\subsection{Comparison with \cite{toba2017}}
\begin{figure}[h]
	\centering
	\includegraphics[width=8cm, bb=0 0 360 360]{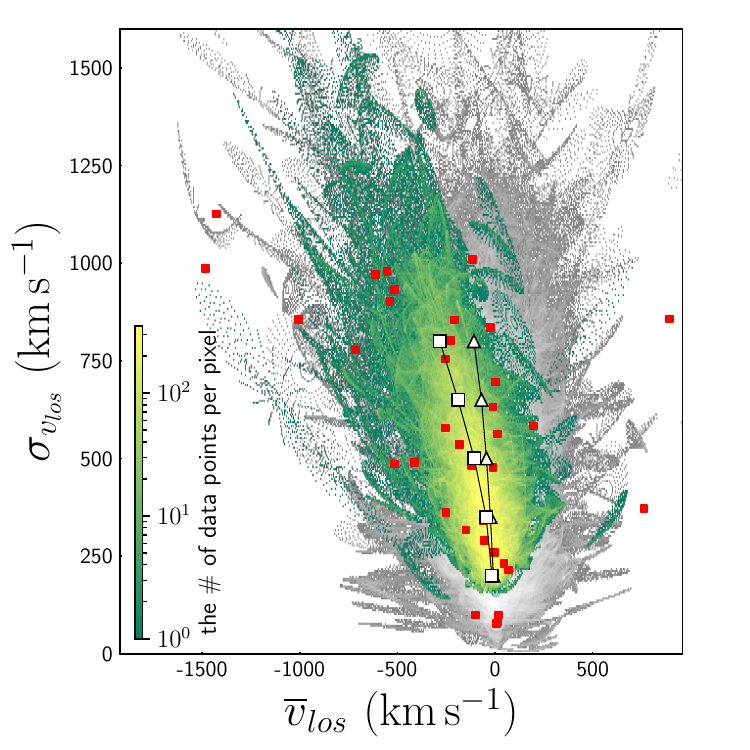}
	\caption{\Add{
			Comparison of the theoretical model of this study with the VVD diagram of \cite{toba2017} (i.e., red markers).
			The colored VVD diagrams for gas fraction of 30\% models and gray VVD diagrams for gas fraction of 10\% models.
			The mean value for \voff of each \sigv bin is indicated by square for gas fraction 30\% and triangle symbols for gas fraction 10\%.}
		\label{fig:vvd_toba17}}
\end{figure}

\Add{\cite{toba2017} reported that some IR bright DOGs with fast \OIII outflows exhibit velocity dispersion and velocity offsets higher than 500 km s$^{-1}$.
We compare the VVD diagrams of our model with \cite{toba2017} in Figure \ref{fig:vvd_toba17}.
Figure \ref{fig:vvd_toba17} shows that our model covers most sources observed in \cite{toba2017}.
However, in \cite{toba2017}, a few objects with \sigv = 1000 km s$^{-1}$ and \voff = -1000 km s$^{-1}$ are observed, and the sources was tilted more strongly than in our model.
The tilt of the VVD diagram was determined by the dust extinction based on the results of this study.
This implies that the high-velocity outflow sources observed in \cite{toba2017} may be even more obscured than those in our results.
This can be reproduced using a galaxy merger model with a higher gas fraction.
We were unable to select DOGs from our simulation data because we did not conduct infrared pseudo-observations in this study.
For comparison with more detailed observations, both infrared and ionizing emission line pseudo-observations should be performed.}\par

\subsection{model limitation}
\Add{We performed identical galactic-core merger simulations. 
In our models, we fixed the initial mass of SMBH at 10$^8M_\odot$.
This is because \cite{toba2017} suggested that the SMBH mass of IR bright DOGs is approximately 10$^8 M_\odot$.
DOGs that contain even heavier SMBHs than 10$^8 M_\odot$ are often observed as Hot DOGs \citep[i.e.,][]{wu2018}.
In addition to IR bright DOGs, observational studies have suggested that Hot DOGs are also accompanied by very fast ionization flows \citep[e.g.,][]{jun2020}.
However, at redshift 2, SMBHs of 10$^9M_\odot$ are considered to be approximately 100 times rarer than SMBH at 10$^8 M_\odot$ \citep[e.g.][]{rosas2016}.
Therefore, we consider that an SMBH mass of 10$^8M_\odot$ is the most appropriate initial condition for a galaxy merger simulation in considering the high-velocity outflows associated with buried AGNs.}\par

\Add{In addition, it is believed that buried AGNs are formed during the late stages of major merger \citep[e.g.][]{dey2009,narayanan2010,yamada2021,yutani2022,dougherty2023}.
Our identical major merger simulation in this study is based on a major merger scenario.
However, IR bright DOGs could be also formed by a minor merger with a galaxy, accompanied by an SMBH of about 10$^7 M_\odot$.
We suspect that as long as the central nuclei are buried in dust, the apparent effect of dust extinction suggested in this study on the observed outflow velocity would also be the case. The cases of BH masses are not idendical and minor mergers will be examined elsewhere.}

\section{summary and conclusions} \label{sec:summary}
We \Add{simulated} of late-stage galaxy mergers using the N-body/SPH code \texttt{ASURA} \citep{saitoh2008,saitoh2009,saitoh2013}.
Our goal \Add{is} to explore the velocity characteristics of ionized outflows during the late stages of galaxy mergers.
Our study \Add{yields} the following key findings:
\begin{enumerate}
	\item Late-stage galaxy mergers exhibit strong ionized outflows with velocity dispersions surpassing 500 km s$^{-1}$. (Figure \ref{fig:vvd_merger}).
	\item \Add{The mean \Adds{observed} velocity (\voff) of the ionized outflows is significantly influenced by dust extinction. 
	In our models, higher gas fractions result in a greater tilt of the VVD diagram toward a blueshift (Figures \ref{fig:vvd_merger} and \ref{fig:vvd_tau}).}
	\item The velocity dispersion (\sigv) of the ionized outflows are also affected by dust extinction. The \OIII emission lines from dense gas are attenuated by dust grains, leading to higher velocity dispersion, enabling the observation of more diffused and hotter gas particles. This trend holds particular importance for buried AGNs. (Figure \ref{fig:vvd_tau})
\end{enumerate}
\Add{Although} the luminosity and equivalent width of emission lines were not discussed in this study, we intend to perform pseudo-observations of ionized emission lines part of a future research.

\begin{acknowledgments}
	We are grateful to Takayuki Saito for providing us with the \texttt{ASURA} code \citep{saitoh2008,saitoh2009,saitoh2013}.
	Numerical computations were performed on a Cray XC50 at the Center for Computational Astrophysics, National Astronomical Observatory of Japan. \Adds{We also thank the anonymous referee for constructive comments on the text.}
\end{acknowledgments}

\software{ASURA \citep{saitoh2008,saitoh2009,saitoh2013}, \Add{AGAMA \citep{vasiliev2019}}}

\appendix

\section{Convergence of resolution}\label{app:convergence}
\begin{figure*}[h]
	\centering
	\includegraphics[width=\linewidth,bb=0 0 1080 360]{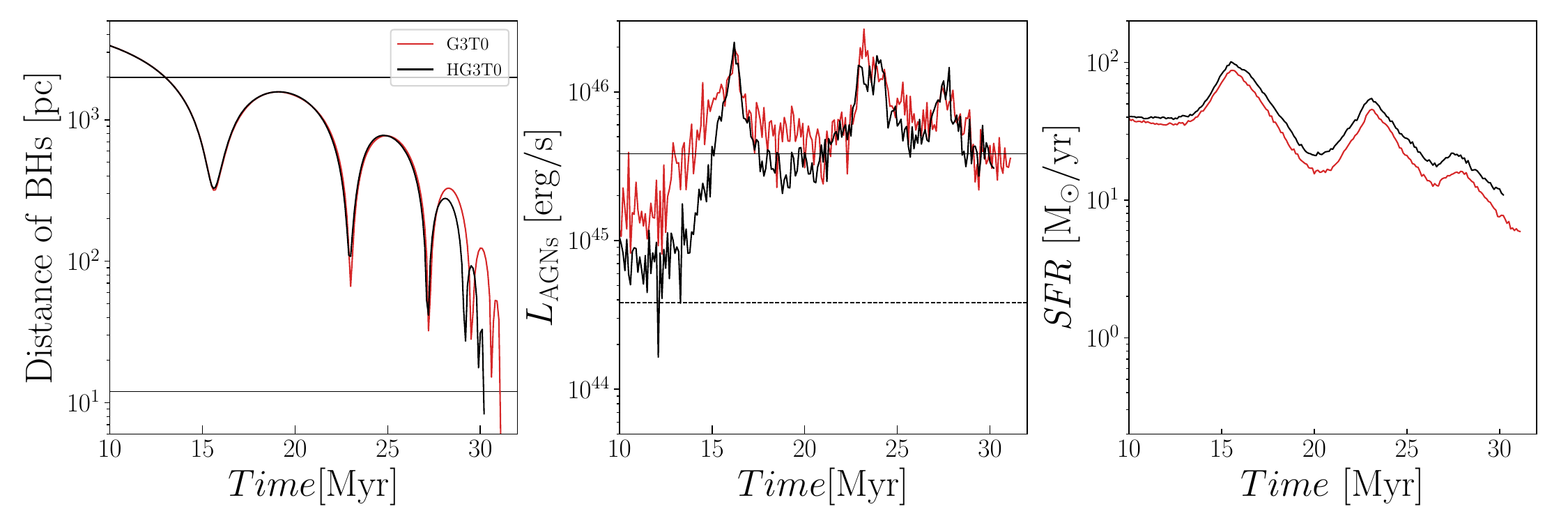}
	\caption{Comparison between G3T0 and HG3T0 with respect to the distance BH-binary (left panel), total AGN luminosity (center panel), and star formation rate (right panel).\label{fig:app1}}
\end{figure*}

We performed galaxy merger simulations with a higher resolution model (here after HG3T0) than \Add{with the} G3T0. 
Table \ref{tab:app_table} reports the model parameters for HG3T0.
The spatial resolution \Add{was} 2 pc, \Add{the} stellar particle mass resolution \Add{was} 1.25$\times 10^4 M_\odot$, and \Add{the} gas particle mass resolution \Add{was} 2$\times$10$^3 M_\odot$ for the HG3T0 model.
Figure \ref{fig:app1} shows the distance between \Add{the} binary BHs, total AGN bolometric luminosity, and star formation rate for the HG3T0 and G3T0 models.
There is no significant difference between the two models. For star formation rates, the results obtained with the HG3T0 model are systematically higher due to its higher spatial resolution; the difference in AGN luminosity before 15 Myr is due to the randomness of the gas clumps formed by the ``fluctuations'' in the initial gas disk.\par

\begin{deluxetable*}{cccccccccccc}[]
	\tablenum{3}
	\tablecaption{Initial parameters of the pre-merger system\label{tab:app_table}}
	\tablewidth{0pt}
	\tablehead{
		Name$^{\rm a}$&${M_{BH}}^{\rm b}$&${M_{bul}}^{\rm c}$& ${M_{\rm gas}}^{\rm d}$&${\epsilon}^{\rm e}$&${r_{acc}}^{\rm f}$&${r_{h}}^{\rm g}$&${R_{\rm gas}}^{\rm h}$&${\Delta M_{bul}}^{\rm i}$&${\Delta M_{\rm gas}}^{\rm j}$&${\mu_{gas}}^{\rm k}$\\
		&[$10^8 M_{\odot}$]&[$10^{11} M_{\odot}$]&[$10^{10} M_{\odot}$]&[pc]&[pc]&[kpc]&[kpc]&[$10^4 M_{\odot}$]&[$10^3 M_{\odot}$]&
	}
	\startdata
	HG3 & 1.0 &  1.0 & 1.0 & 2.0 & 4.0 & 2.2 & 1.0 & 1.25 & 2.0 & 30\%\\
	\enddata
	\tablecomments{
		(a) Name of the pre-merger system.
		(b) Mass of sink, (c) star, and (d) gas particles, respectively.
		(e) Gravitational softening
		(f) Accretion radius of sink particle.
		(g) Effective radius \Add{of} stellar bulge Sersic profile.
		(h) Outer edge radius of uniform-density gas disk.
		(i) Mass resolution of star particles, (j) and gas particles.
		(k) Gas fraction within 1kpc.}
\end{deluxetable*}

\bibliography{yutani2023}{}
\bibliographystyle{aasjournal}



\end{document}